\documentclass[12pt]{article} 
\usepackage{amssymb}
\usepackage{authblk}
\usepackage{amsmath} \usepackage{amsthm} \usepackage{latexsym}
\usepackage{graphicx} \usepackage{layout} \usepackage{enumerate}
\usepackage{amsfonts}
\usepackage{amssymb}
\usepackage{amsmath}
\usepackage{bbm}
\usepackage{color}

\newcommand{\nada}[1] {} \newcommand{\M} {\mathbb M} \newcommand{\Div}
{{\rm div}}     \newcommand{\E} {\ensuremath{\mathbb E}}
\newcommand{\R} {\ensuremath{\mathbb R}} \newcommand{\N}
{\ensuremath{\mathbb N}} \newcommand{\T} {\ensuremath{\mathbb T}}
\newcommand{\Z} {\ensuremath{\mathbb Z}}  
  \newcommand{\eps} {\epsilon} 
 \newcommand{\grad} {\nabla}

\newcommand{\df} {\,{\rm d}}

\renewcommand{\S}{\mathcal{S}}
\newcommand{\I}{\mathcal{I}}
\newcommand{\norm}[1]{\left\lVert#1\right\rVert}
\newcommand{\abs}[1]{\left|#1\right|}
\newcommand{\ip}[2]{\langle#1,#2\rangle}
\newcommand{\1}{\mathbbm{1}}
\newcommand{\lra}{\longrightarrow}
\newcommand{\ra}{\rightarrow}
\newcommand{\lin}{\mathrm{lin}}
\newcommand{\g}{\gamma}
\title{Entropic and gradient flow formulations for nonlinear diffusion}
\author[1]{Nicolas Dirr\thanks{DirrNP@cardiff.ac.uk}}
\author[2]{Marios Stamatakis\thanks{M.G.Stamatakis@bath.ac.uk}}
\author[2]{Johannes Zimmer\thanks{zimmer@maths.bath.ac.uk}} 
\affil[1]{Cardiff University, School of Mathematics, Senghennydd Road, Cardiff CF24 4AG, UK}
\affil[2]{University of Bath, Department of Mathematical Sciences, Bath BA2 7AY, UK}

\begin{document}

\maketitle

\begin{abstract}
  Nonlinear diffusion $\partial_t \rho = \Delta(\Phi(\rho))$ is considered for a class of nonlinearities $\Phi$. It is
  shown that for suitable choices of $\Phi$, an associated Lyapunov functional can be interpreted as thermodynamic
  entropy. This information is used to derive an associated metric, here called thermodynamic metric. The analysis is
  confined to nonlinear diffusion obtainable as hydrodynamic limit of a zero range process. The thermodynamic setting
  is linked to a large deviation principle for the underlying zero range process and the corresponding equation of
  fluctuating hydrodynamics. For the latter connections, the thermodynamic metric plays a central role.
\end{abstract}


\section{Introduction}
There is a wealth of results on nonlinear diffusion equations of the kind
\begin{equation}
  \label{eq:nonlin}
  \partial_t \rho = \Delta(\Phi(\rho));
\end{equation}
as the monumental monograph by Vazquez~\cite{Vazquez2007a} for the porous medium equation, $\Phi(\rho) = \rho^m$,
demonstrates. Recently, Bodineau, Lebowitz, Mouhot and Villani~\cite{Bodineau2014a} studied drift-driven nonlinear
diffusion equations with inhomogeneous Dirichlet data and established the existence of Lyapunov functionals for such
equations.

Starting with the work of Jordan, Kinderlehrer and Otto~\cite{Jordan1998a}, diffusion equations have been written as
gradient flow of the entropy. For the example of linear diffusion, $\partial_t \rho = \Delta \rho$, with the Boltzmann
entropy density $S_\lin := \rho \log(\rho)$, it is immediate that
$\partial_t \rho = \mathrm{div} (\rho \nabla S_\lin')$. The seminal insight of~\cite{Jordan1998a} is that the operator
involved, $K(\rho)\xi := - \mathrm{div} (\rho \nabla \xi)$, is associated to a metric and hence defines a geometry; the
evolution is then determined by the steepest descent of the entropy $\S$ in this associated geometry. Here $\S$ is the
classic entropy given by $\S(\rho)=\int S_\lin(\rho(x))dx$ if $\rho$ is absolutely continuous with respect to the
Lebesgue measure and $\S(\rho)=+\infty$ if $\rho$ is not absolutely continuous. More generally, gradient flows are
formally equations of the form $\partial_t \rho = - K(\rho) d\S(\rho)$, where $\S$ is a functional defined on the state
space, $d\S$ the Fr\'{e}chet differential and $K(\rho)$ an operator from the cotangent space to the tangent, the
so-called Onsager operator. This formulation says more than that $\S$ is a Lyapunov functional; it says the evolution
is given by the steepest descent evolution of $\S$ in the geometry given by $K$.

In this short note, we ask a simple question: For equation~\eqref{eq:nonlin}, with a suitable choice of $\Phi$, can one
show that an associate Lyapunov functional is in fact a thermodynamic entropy, that is, the entropy associated with an
underlying particle system? This requires that~\eqref{eq:nonlin} is the hydrodynamic limit of the underlying particle
system. Furthermore, we will then write~\eqref{eq:nonlin} as formal gradient flow (steepest descent) of the entropy, in
a metric we derive as the metric associated with the thermodynamic entropy. We thus call this metric (or geometry) the
thermodynamic metric (or geometry).  In some cases, the formal setting can be shown to define rigorously a gradient
flow.

We employ basic tools, techniques and results from statistical mechanics and optimal transport. Yet, the result
seems interesting in a twofold way. Firstly, from the statistical mechanics perspective, the thermodynamic entropy is
well known (see, e.g., Grosskinsky~\cite{Grosskinsky2004a}), but has not been placed in a gradient flow context,
i.e., linked to the associated thermodynamic metric. Secondly, the nonlinear diffusion
$\partial_t\rho=\Delta\Phi(\rho)$ has been studied in a classic Wasserstein setting, with the $2$-Wasserstein
metric~\cite{Ambrosio2005a}, and has been shown to be the gradient flow, with respect to the $2$-Wasserstein metric, of
the free energy $\mathcal{F}$ given by
\begin{equation*}
  \mathcal F(\rho)=\int F(\rho)\df x+F'(\infty)\rho^\perp(\T^d) . 
\end{equation*}
Here $F$ is characterised by the ordinary differential equation $\Phi(\rho)=\rho F'(\rho)+F(\rho)$ and assumed to
satisfy $F'(\infty):=\sup_{t>0}\frac{F(t)}{t} =+\infty$ (see~\cite[Theorem 11.2.5]{Ambrosio2005a} for less restrictive
assumptions). This gradient flow setting is different from the setting presented in this article.

It is well known that one equation can have different incarnations as gradient flow, that is, as steepest descent of a
functional in a metric space; linear diffusion is a classic example with infinitely many such formulations. Then the
functional is a Lyapunov functional. Here we show that in the setting of this article, the Lyapunov function has a
thermodynamic interpretation, being the thermodynamic entropy of the so-called zero range process. We point out that
the existence theory can be established without this thermodynamic interpretation. That is, the knowledge of Lyapunov
functionals, which may or may not be entropies, can be key to a successful existence theory, as~\cite{Fornaro2012a}
testifies. At the same time, large deviation theory singles out one particular metric in the cases under
consideration. We show in Section~\ref{sec:Connection-to-large} that this gives, at least formally, immediately rise to
a variational scheme of Jordan-Kinderlehrer-Otto (JKO) type~\cite{Jordan1998a}. 

We remark that different particle processes can lead to the same equation as hydrodynamic limit, but that equation will
then typically have different representations as pairs of metric and entropy. For example, the symmetric exclusion
process and a system of independent random walks both lead to the linear diffusion equation, but with two different
entropies and correspondingly two different metrics, see~\cite{Adams2013a}.

It turns out that in the setting of this article, the same pair of a weighted Wasserstein metric and the thermodynamic
entropy arises both from large deviation theory and a weighted version of the classical Benamou-Brenier formula. The
weighted Wasserstein metric is (formally) conformally equivalent to the classical one, with the weight being related to
the diffusion of a tagged particle. Namely, it is often convenient to write~\eqref{eq:nonlin} as
\begin{equation}
  \label{eq:nonlin-sigma}
   \partial_t\rho=\Delta (\sigma(\rho)\rho) =\Div (\sigma(\rho)\grad \rho)+\Div (\rho\grad\sigma(\rho)),
\end{equation}
since $\sigma(\rho) = \Phi(\rho)/\rho$ will be related to the diffusivity of a tagged particle in the underlying zero
range process, as discussed in Subsection~\ref{sec:Asympt-behav-tagg}. Roughly speaking, the equation which is the
hydrodynamic limit of the zero range process looks like the Kolmogorov equation of a single tagged particle. This
diffusivity (sometimes called {\em self-diffusion coefficient}) will turn out to be closely related to the
\emph{thermodynamic metric} of the process discussed below.

\paragraph{Plan of the paper} We focus on the so-called zero-range process as underlying microscopic model; this
process and its properties are described in Section~\ref{sec:Particle-models}. In particular, the entropy (density) $S$
is identified there; in Section~\ref{sec:Therm-form-nonl} the associated metric is determined. We show that this metric
can also be obtained differently, by a large deviation argument; this is the content of
Section~\ref{sec:Connection-to-large}, which also sketches connections to associated models of fluctuating
hydrodynamics.

\section{Particle models}
\label{sec:Particle-models}

\subsection{The zero range process}
\label{sec:zero-range-process}

In this subsection, we briefly summarise the zero range process on a flat torus;
see~\cite{Grosskinsky2004a,Grosskinsky2003a} for more information on zero range process and the condensation phenomenon
they exhibit for particular choices of jump rates.

A zero range process is an interacting particle system. Let $\Lambda$ be lattice, that is, an ordered collection of
boxes indexed by an (at most countable) index set. Each box can be empty or contain a finite number of particles. The
particles are indistinguishable. They evolve according to an irreducible Markov kernel on $\Lambda$ in the following
way. A particle at a given site $x\in\Lambda$ interacts only with the particles located at the same site (hence zero
range). Specifically, the rate at which one particle leaves a site $x$ depends only on the total number of particles at
site $x$, described by the (local) \emph{jump rate function} $g\colon\N_0 := \N \cup \{0\} \lra\R^+_0$, which satisfies
$g(k)=0$ if and only if $k=0$. Here $g$ is assumed to be Lipschitz continuous, i.e.,
$\sup_{k\in\N_0}|g(k+1)-g(k)|<+\infty$.

In particular, for the hydrodynamic limit in the flat torus $\T^d$ we consider the diffusively rescaled zero range
processes with finite lattice the discrete $d$-dimensional torus $\T_N^d\cong(\,^\Z/_{N\Z})^d\cong\{0,1,\dots,N-1\}^d$
as $N\rightarrow+\infty$. The state space of the system is the set $\M_N:=\N_0^{\mathbbm{\T}_N^d}$ of all
configurations of particles $\eta\colon\mathbbm{T}_N^d\longrightarrow\N_0$; the system is in state $\eta$ if $x$
contains $\eta(x)$ particles for all $x\in\mathbbm{T}_N^d$. Let $p$ denote the symmetric nearest neighbour transition
probability (renormalised so that it has total mass equal to $2d$)
\begin{equation*}
 p(x,y):=\sum_{j=1}^d\1_{\{-e_j,e_j\}}(y-x).  
\end{equation*}
The symmetric (nearest neighbour) zero range process with jump rate $g$ is the Markov jump process on the state space
$\M_N$ with formal generator
\begin{equation*}
  L_Nf(\eta)=N^2\sum_{x,y\in\T_N^d}\{f(\eta^{x,y})-f(\eta)\}g(\eta(x))p(y-x),
\end{equation*}
where $f\colon\M_N\lra\R$ is a function on the space of configurations, $\eta^{x,y}:=\eta-\1_{\{x\}}+\1_{\{y\}}$ is the
configuration resulting from $\eta$ by moving a particle from $x$ to $y$ and the term $N^2$ comes from the diffusive
rescaling of the process.

The microscopic stochastic dynamics defined by this generator can be described as follows. Starting from an initial
configuration $\eta_0$, the system waits an exponential time $\tau_1$ with rate
$N^2\lambda(\eta_0):=N^2\sum_{x\in\mathbbm{T}_N^d}g(\eta_0(x))$, at which time one particle is moved from $x$ to $y$ with
probability $\frac{g(\eta_0(x))}{\lambda(\eta_0)}p(y-x)$. Then the processes starts again with
$\eta_{\tau_1}:=\eta_0^{x,y}$ in place of $\eta_0$.

The equilibrium states of zero range processes are well known. They are characterised by the equation
$\nu L_N=0$~\cite[Prop. A.4.1]{Kipnis1999a}. If $\nu$ is a translation invariant measure on the configuration space
$\mathbbm{M}_N$, all its one-site marginals $\nu_x(\cdot):=\nu\{\eta(x)= \cdot\}$, $x\in\T_N^d$, must be equal. In
particular, $\varphi:=\E_\nu g(\eta(0))=\E_\nu g(\eta(x))$. So if $\nu$ is a translation invariant and product
equilibrium state with \emph{fugacity} $\varphi:=\E_\nu g(\eta(0))$ , then $\nu L_N=0$ can be solved to yield
\begin{equation*}
  \nu\{\eta \bigm| \eta(0)=k\}=\nu\{\eta \bigm| \eta(0)=0\}\cdot\frac{\varphi^k}{g!(k)},
  \quad (x,k)\in\mathbbm{T}_N^d\times\N_0 
\end{equation*}
where $g!(k):=g(1)\cdot\ldots\cdot g(k)$ and $g!(0) = 1$. Since $\nu$ is a probability measure, the \emph{partition
  function}
\begin{equation*}
  Z(\varphi):=\sum_{k=0}^\infty\frac{\varphi^k}{g!(k)}
\end{equation*}
converges at $\varphi$ and $\nu\{\eta\bigm|\eta(0)=0\}=\frac{1}{Z(\varphi)}$. Conversely, for $\varphi$ in the proper
domain $D_Z:=\{\varphi\in\R^+_0 \bigm| Z(\varphi)<+\infty\}$, the translation invariant product measure
$\bar{\nu}_{\varphi}^N$ in $\M_N$ with one-site marginal
\begin{equation*}
  \bar{\nu}_\varphi^N\{\eta(0)=k\}:=\frac{1}{Z(\varphi)}\frac{\varphi^k}{g!(k)}
\end{equation*}
is an equilibrium state of the zero range process. Consequently the family of the translation invariant and product
equilibrium states of the zero range process is $\{\bar{\nu}_\varphi^N\}_{\varphi\in D_Z}$. From a statistical
mechanics point of view, this family is the grand canonical ensemble. Of course, in order for non-trivial equilibrium
states to exist, the critical fugacity
\begin{equation}
  \label{DefCritFug}
  \varphi_c:=\liminf_{k\ra+\infty}\sqrt[k]{g!(k)}
\end{equation}
must be strictly positive to ensure that the partition function $Z$ has non-trivial radius of convergence.

For the description of the evolution of the empirical density of the zero range process, it is useful to have a
reparametrisation of the grand canonical ensemble by the density. This can be achieved by introducing the \emph{density
  function} $R\colon D_Z\lra[0,+\infty]$, which is given by
\begin{equation}
  \label{eq:DensityFunction} 
  R(\varphi):=\E_{\bar{\nu}_{\varphi}^N}\eta(0)=\frac{\varphi}{Z(\varphi)}
  \sum_{k=0}^\infty k\frac{\varphi^{k-1}}{g!(k)}=\frac{\varphi Z'(\varphi)}{Z(\varphi)}.
\end{equation}
This function is analytic (on $[0,\varphi_c)$) and strictly increasing~\cite[Section 2.3]{Kipnis1999a} with
$D_R=D_Z\cap D_{Z'}$.  The critical density is then $\rho_c:=\sup R(D_R)\in(0,+\infty]$. In this article, we restrict
our attention to zero range processes with superlinear jump rate $g$, i.e., for some constant $a_0>0$
\begin{equation}
  \label{ineq:superlinjumprate}
  g(k)\geq a_0k \text{ for all }k\geq 1,
\end{equation}
which enforces $\varphi_c=\rho_c=+\infty$ (we refer the reader to~\cite{Grosskinsky2003a} for zero range processes that
exhibit phase transitions). Then the inverse $\Phi:=R^{-1}$ is well defined on all of $\R^+_0$; by setting
$\nu_\rho^N:=\bar{\nu}^N_{\Phi(\rho)}$, the grand canonical ensemble is reparametrised by the density. We note that
$\Phi(\rho)$ is stochastically represented as the mean local jump rate with respect to the translation invariant and
product equilibrium state $\nu_\rho^N$ of density $\rho$, since
\begin{equation}
  \label{eq:MeanJRFunction}
  \E_{\nu_{\rho}^N}g(\eta(0))=\frac{1}{Z(\Phi(\rho))}\sum_{k=1}^\infty g(k)\frac{\Phi(\rho)^k}{g!(k)}=\Phi(\rho).
\end{equation} 

The \emph{thermodynamic entropy density} $S$ of the zero range process is defined as the Legendre transform of the
pressure $P(\lambda) = \log(Z(\exp(\lambda))$; here it is (see~\cite[Eq. (3.2)]{Grosskinsky2003a}; we use the opposite
sign convention)
\begin{equation}
  \label{eq:thermo-ent}
  S(\rho) = \rho \log\Phi(\rho) - \log(Z(\Phi(\rho)))
\end{equation}
and thus
\begin{equation}
  \label{eq:thermo-ent-diff}
  S'(\rho) = \log(\Phi(\rho)) + \rho \frac{\Phi'(\rho)}{\Phi(\rho)} - \frac{Z'(\Phi(\rho))}{Z(\Phi(\rho))} \Phi'(\rho).
\end{equation}
Since $\Phi=R^{-1}$, inserting $\varphi=\Phi(\rho)$ in~\eqref{eq:DensityFunction} we get
\begin{equation*}
  \rho=R(\Phi(\rho))=\frac{Z'(\Phi(\rho))}{Z(\Phi(\rho))}\Phi(\rho) ,
\end{equation*}
and so
\begin{equation*}
\frac{Z'(\Phi(\rho))}{Z(\Phi(\rho))} \Phi'(\rho)=\rho\frac{\Phi '(\rho)}{\Phi(\rho)},
\end{equation*}
and hence~\eqref{eq:thermo-ent-diff} simplifies to
\begin{equation}
  \label{eq:thermo-ent-diff-simp}
  S'(\rho) = \log(\Phi(\rho))
    = \log(\sigma(\rho)\rho).
\end{equation}

The large deviations functional with speed $\frac{1}{N^d}$ (see also Subsection~\ref{sec:Large-devi-theory}) of the
occupation variables $\{\eta(x)\}_{x\in\Z^d}$ with respect to the equilibrium state $\nu_{\rho_*}^\infty$ on the
infinite lattice $\Z^d$ is given by~\cite[Lemma 6.1.7]{Kipnis1999a}
\begin{equation*}
	I_{\rho_*}(\rho)=\rho\log\frac{\Phi(\rho)}{\Phi(\rho_*)} -\log\frac{Z(\Phi(\rho))}{Z(\Phi(\rho_*))}.
\end{equation*}
Thus it differs from the thermodynamic entropy by an affine function that depends only on the choice of the equilibrium
state $\nu_{\rho_*}^N$ through its density $\rho_*$.

The integral functional $\S$ corresponding to the thermodynamic entropy $S$ is given by
\begin{equation}
  \label{Def:ThermEntrFunct}
  \S(\rho)=\int_{\T^d}S(\rho^{ac})\df x+S'(\infty)\rho^\perp(\T^d),
\end{equation}
where $S'(\infty)=\lim_{\rho\ra+\infty}S(\rho)/\rho=\log\varphi_c$ is the rate of linear growth at infinity of the
thermodynamic entropy, and $\rho=\rho^{ac}+\rho^\perp$ is the Radon-Nikodym decomposition of $\rho$,
$\rho^{ac}\ll\mathcal{L}_{\T^d}$, $\rho^\perp\perp\mathcal{L}_{\T^d}$ (with respect to Lebesgue measure
$\mathcal{L}_{\T^d}$ on the torus). In particular, in the case that $\varphi_c=+\infty$, the thermodynamic entropy has
superlinear growth $S'(\infty)=+\infty$, and thus $\S(\rho)=+\infty$ if $\rho\not\ll\mathcal{L}_{\T^d}$. The integral
functional $\I_{\rho_*}$ associated to the rate functional $I_{\rho_*}$ as in~\eqref{Def:ThermEntrFunct},
\begin{equation*}
  \mathcal{I}_{\rho_*}(\rho)=\int_{\T^d}I_{\rho_*}(\rho^{ac})\df x+I_{\rho_*}'(\infty)\rho^\perp(\T^d),
\end{equation*}
is (see the proof of Lemma 5.1.6 in~\cite{Kipnis1999a}) the rate functional for the large deviations principle (with
speed $\frac{1}{N^d}$) satisfied by the sequence
\begin{equation*}
  \Big(\frac{1}{N^d}\sum_{x\in\T_N^d}\eta(x)\delta_{x/N}\Big)_\sharp\nu_{\rho_*}^N
  \in\mathcal{P}(\mathcal{M}_+(\T^d)),\quad N\in\N,
\end{equation*}
of the empirical embeddings of the zero range process with respect to an equilibrium state of density $\rho_*$. In the
last display $\mathcal{M}_+(\T^d)$ denotes the set of all non-negative Borel measures on the torus, and $\mathcal{P}$
the set of probability measures; the push-forward $f_\sharp \mu$ of a measure $\mu$ by a (measurable) map $f$ is
defined as the measure satisfying $f_\sharp \mu (A) = \mu(f^{-1}(A))$ for all (measurable) sets $A$. Since $S$ and
$I_{\rho_*}$ differ by an affine function, the integral functionals $\S$ and $\I_{\rho_*}$ differ by a constant and
consequently $\S$ and $\I_{\rho_*}$, $\rho_*\in\R^+_0$ define the same gradient flow (with respect to any metric).

It is well known (see~\cite[Theorem 5.1.1]{Kipnis1999a} for a precise formulation) that the hydrodynamic limit of the
zero range process is given by~\eqref{eq:nonlin}, with $\Phi$ being the mean jump rate function. This means that
starting the process from a sequence $\{\mu_0^N\}$ of initial distributions associated to a profile
$\rho_0\in L^1(\T^d)$, i.e., from a sequence of initial laws $\mu_0^N$, $N\geq 1$, such that for all $\delta>0$ and
$G\in C(\T^d)$
\begin{equation*}
  \lim_{N\ra+\infty}\mu_0^N\left\{\abs{\frac{1}{N^d}\sum_{x\in\T_N^d}G\left(\frac{x}{N}\right)\eta(x)
  -\int_{\T^d}G(x)\rho_0(x)\df x}>\delta\right\}=0,
\end{equation*}
then the law $\mu_t^N$ of the process at each time $t>0$ is associated to the profile $\rho_t$, where
$(\rho_t)_{t\in\R^+_0}$ is the unique weak solution of~\eqref{eq:nonlin} starting from $\rho_0$. Furthermore, if the
initial profile is of class $C^{2+\theta}$ for some $\theta>0$, then the relative entropy method shows that
asymptotically as $N\ra+\infty$ the law $\mu_t^N$ looks like the product measure $\nu_{\rho_t(\cdot)}^N$ with slowly
varying profile $\rho_t$, which is characterised by the marginals
\begin{equation*}
  \nu_{\rho_t(\cdot)}^N\{\eta(x)=k\}=\frac{1}{Z(\Phi(\rho_t(x/N)))}\frac{\Phi(\rho_t(x/N))^k}{g!(k)}; 
\end{equation*}
see~\cite[Theorem 6.1.1]{Kipnis1999a} for a precise formulation.

\subsection{Asymptotic behaviour of tagged particles}
\label{sec:Asympt-behav-tagg}

Given a symmetric zero-range process, we can ``tag'' a particle that is originally at the origin, say, and follow its
movements, denoting its position at time $t$ by $X(t).$ It is natural to ask whether there is a central limit theorem
for this tagged particle, that is, whether the law of $N^{-1}X(N^2t)$ converges to a diffusion $dX=\sigma(X)dW$, where
$W$ is a standard Brownian motion, and if so, what the diffusivity $\sigma$ is. This central limit theorem has been
shown to hold in many situations, both in equilibrium (i.e., with the initial distribution of the particles given by an
invariant measure for the process) and in non-equilibrium, see, e.g.,~\cite{Kipnis1999a} and the references therein. It
turns out~\cite{Jara2013a} that for $\rho>0$,
\begin{equation*}
  \sigma(\rho) =\frac{\Phi(\rho)}{\rho},
\end{equation*}
where $\Phi$ and $\rho(t,x)$ as in~\eqref{eq:nonlin}. In the next section, this intrinsic diffusivity will be related
to what we will call below the thermodynamic metric of the process.

\section{Thermodynamic formulation}
\label{sec:Therm-form-nonl}

We now phrase~\eqref{eq:nonlin-sigma} with nonlinear $\sigma$ in a Wasserstein setting. We extend the classic
Benamou-Brenier formulation~\cite{Benamou2000a} of optimal transport by a weighted version.

As a starting point, note that with the notation from the previous section (see, e.g.,~\eqref{eq:thermo-ent-diff-simp})
\begin{equation}
  \label{eq:nonl-ww}
  \partial_t\rho=\Delta(\Phi(\rho))={\rm div}(\Phi(\rho)\grad S^\prime(\rho))={\rm div}(\rho \sigma \grad S^\prime(\rho))
\end{equation}
with $\sigma(\rho)=\Phi(\rho)/\rho$. In order to simplify notation, we will often omit the dependence of $\sigma$ on
$\rho$.

In this section we will show that this structure allows us to interpret~\eqref{eq:nonlin-sigma} as a gradient flow of
the entropy~\eqref{Def:ThermEntrFunct} with respect to a metric which is associated to the operator
${\rm div} (\Phi(\rho)\nabla\cdot )$ appearing in the equation above. This metric will be shown to be a weighted
Wasserstein metric with weight $\sigma^{-1}.$

Here we will assume that our spatial domain is the $d$-dimensional flat torus $\T^d$. For general bounded domains and
homogeneous Dirichlet boundary data the reasoning is analogous. The case of inhomogeneous boundary data is covered in
\cite{Bodineau2014a}.

We introduce a metric tensor, following the work of Otto~\cite{Otto2001a}. See also the presentation in~\cite[Section
8]{Villani2003a}. Here we work with the set $\mathcal{P}_{ar}\T^d$ of probability measures on the torus with continuous
Lebesgue densities.  What follows can be seen as the Benamou-Brenier formulation of optimal transport with a
nonconstant metric. In this context, a point $\rho\in\mathcal{P}_{ar}\T^d$ in this set has formally as tangent space
the space of all infinitesimal variations at $t=0$ of smooth curves $(-\epsilon,\epsilon)\ni t\mapsto\rho_t$ with
$\rho_0=\rho$. By the conservation of mass, these variations satisfy $\int \partial_t|_{t=0}\rho_t \df x=0$ and thus
the tangent space as a set can be identified with the set of all continuous functions $\zeta\colon \T^d\lra\R$ such
that $\int\zeta \df x=0$. The tangent spaces are thus all equal as sets, while the metric we are going to define
depends on the base point $\rho$. As in the standard case (e.g.,~~\cite{Otto2001a}), a metric is defined on each
tangent space by associating to each tangent vector $\zeta=\partial_t|_{t=0}\rho_t$ (i.e., a derivative of a path in
the space of densities/measures) a vector field representing this variation, via the equation
\begin{equation}
  \label{eq:cont}
  \zeta + \Div (\rho v) = 0,
\end{equation} 
with unknown the vector field $v\colon \T^d\to\R^d$. To ensure uniqueness of the vector field $v$ representing the
tangent vector $\zeta$, we minimise the \emph{weighted} kinetic energy under the constraint~\eqref{eq:cont},
\begin{equation}
  \label{eq:kinen}
  \inf_v \left\{ \int_{\T^d} \rho \sigma^{-1} \abs{v}^2 \df x \bigm| \zeta  + \Div (\rho v) = 0 \right\}. 
\end{equation}
In the classic case (no division by $\sigma$), this term has an interpretation as twice the kinetic energy, and we thus
call~\eqref{eq:kinen} (twice) the \emph{weighted kinetic energy}.  We take outer variations of the form
$v+\eps \rho^{-1}w$ with divergence-free $w$; note that the latter property ensures that~\eqref{eq:cont} is
preserved. Then the extremality condition for~\eqref{eq:kinen} reads
\begin{equation*}
  \int_{\T^d} \sigma^{-1} \ip v w \df x =0
\end{equation*}
for all divergence-free $w$, where $\ip \cdot\cdot$ is the usual Euclidean scalar product on $\R^d$. By the
Helmholtz-Hodge decomposition of vector fields into divergence-free and gradient fields, this implies
\begin{equation}
  \label{eq:eta}
  \sigma^{-1} v=\grad \xi
\end{equation}
for some real-valued function $\xi$, or equivalently $v=\sigma \grad \xi$. Note that thus
$\zeta=-\Div(\rho v)=-\Div(\sigma\rho \grad \xi)$. As (for given suitable $\zeta$) the equation
$-\Div (\rho \sigma \grad \xi )=\zeta$ has a unique solution, there is (up to additive constants) only one such
$\xi=\xi_\zeta$, where the latter notation indicates the dependence on $\zeta$. In this way, we can associate to each
tangent vector $\zeta$ the vector field $v_\zeta:=\sigma(\rho)\nabla\xi_\zeta$. Then on the tangent space at $\rho$, we
define the metric
\begin{equation}
  \label{eq:metric} g_\rho(\zeta_1,\zeta_2):=\int_{\T^d}\frac{\rho}{\sigma(\rho)}\ip {v_{\zeta_1}} {v_{\zeta_2}} 
  \df x=\int_{\T^d}\rho\sigma(\rho)\ip{\nabla\xi_{\zeta_1}}{\nabla\xi_{\zeta_2}}\df x. 
\end{equation}
For each $\rho$, one can think of $\sigma^{-1}(\rho)\ip v v$ as a Riemannian metric on a manifold which is
topologically $\R^d$. This penalises transport in regions with slow diffusion, where $\sigma(\rho)$ is small. In the
classic setting~\cite{Otto2001a}, the division by $\sigma$ is absent. As in classic formal Otto calculus, the cotangent
space is a space of functions modulo constants and thus the function $\xi_\zeta$ corresponding to a tangent vector
$\zeta$ via~\eqref{eq:eta} is the cotangent vector corresponding to $\zeta$ via the metric $g$.
 
Let now $\S$ be any functional of the form~\eqref{Def:ThermEntrFunct} for some integrand $S\colon\R^+_0\lra\R$, which,
by the regularity assumptions made earlier on the probability measures under consideration, is in fact given by
\begin{equation}
  \label{eq:entropy}
  \S(\rho):=\int_\Omega S(\rho) \df x . 
\end{equation} 
We recall that in the usual $L^2$ pairing, the differential $d\mathcal{S}$ is represented by the variational derivative
$\frac{\delta \mathcal{S}(\rho)}{\delta\rho}=S'(\rho)$. Analogously, if we denote by $\langle\cdot|\cdot\rangle_\rho$
the dual pairing between tangent and cotangent space at $\rho$, and if $\zeta$ is represented by
$v_\zeta=\sigma\nabla\xi_\zeta$, an integration by parts yields
\begin{align}
  \label{eq:pair}
  \langle d\S(\rho)|\zeta\rangle_\rho&=\int\zeta\frac{\delta\mathcal{S}(\rho)}{\delta\rho}\df x
                                       =-\int\Div(\rho v_\zeta)S'(\rho)\df x
                                       \nonumber\\
                                     &=\int\rho\ip{v_\zeta}{\nabla S'(\rho)}\df x=\int\frac{\rho}{\sigma(\rho)} \ip{v_\zeta}{{\sigma(\rho)\nabla S'(\rho)}}\df x\nonumber\\
                                     &= g_\rho\big(\zeta,-\Div(\rho\sigma\nabla S'(\rho))\big).
\end{align}
The last equality follows by definition of the metric $g$, since if $\bar{\zeta}:=-\Div(\rho\sigma\nabla S'(\rho))$
then $v_{\bar{\zeta}}=\sigma(\rho)\nabla S'(\rho)$. In words, formula~\eqref{eq:pair} says that the cotangent vector
$d\S(\rho)$ is represented via the metric tensor $g$ by the tangent vector $-\Div(\sigma\rho\nabla S'(\rho))$. Thus in
our case the Onsager operator $K=K_\sigma$ corresponding to the weighted metric $g$ maps $d\mathcal{S}$ to the tangent
vector $\zeta=-\Div(\sigma\rho\nabla[S'(\rho)])$, that is
\begin{equation*}
  K(\rho)d\S(\rho)=-\Div(\sigma\rho\nabla S'(\rho)).
\end{equation*}
Note that the identification $\zeta\mapsto \xi_\zeta$ given via the minimisation of~\eqref{eq:kinen} and~\eqref{eq:eta}
is in fact the inverse of the Onsager operator $K$.

A gradient flow with respect to a metric is distinguished by the fact that the velocity $\partial_t\rho_t$ of a path
$t\mapsto \rho_t$ is related via the metric to the rate of change of a functional, i.e., here
$\zeta=-\Div(\sigma\rho\nabla S'(\rho))$, and so the gradient flow equation for a functional of the
form~\eqref{Def:ThermEntrFunct} becomes
\begin{equation*}
  \partial_t\rho_t=-K(\rho)d\S(\rho)=\Div(\sigma\rho\nabla S'(\rho)).
\end{equation*}
So now if we choose the thermodynamic entropy~\eqref{eq:thermo-ent} as the integrand in the functional $\S$, with its
derivative given by~\eqref{eq:thermo-ent-diff}, then we recover~\eqref{eq:nonlin-sigma}. In summary if we choose the
thermodynamic entropy, then precisely the choice~\eqref{eq:metric} for the metric recovers the nonlinear diffusion
equation~\eqref{eq:nonlin}; we thus call~\eqref{eq:metric} the \emph{thermodynamic metric}.

Weighted kinetic energies as~\eqref{eq:kinen} have been introduced and studied before, in particular by Dolbeault,
Nazaret and Savar\'e~\cite{Dolbeault2009a}, and Carrillo, Lisini and Slep{\v{c}}ev~\cite{Carrillo2010a}. We sketch how
their results can be used to obtain existence and uniqueness for solutions to the gradient flow corresponding
to~\eqref{eq:nonlin} in the thermodynamic setting. The global metric corresponding to the kinetic
energy~\eqref{eq:kinen} is given by
\begin{equation*}
  W_{\sigma;2}^2(\mu,\nu):=\inf \int_0^1\int_{\T^d} \rho \sigma^{-1} \abs{v}^2 \df x  \df t,
\end{equation*}
where the infimum is taken over all pairs $(\rho_t,v_t)_{0\leq t\leq 1}$ with $\rho_0=\mu$, $\rho_1=\nu$ that satisfy
the usual continuity equation $\partial_t\rho + \Div (\rho v) = 0$. Equivalently, one can consider
\begin{equation*}
  W_{\sigma;2}^2(\mu,\nu):=\inf \int_0^1  \int_{\T^d} \Phi(\rho) \abs{v}^2 \df x  \df t
\end{equation*}
with the constraint $\partial_t\rho + \Div (\Phi(\rho) v) = 0$; the latter variational formulation
is~\cite[Eq. (1.6)]{Dolbeault2009a} in our notation. Then, as proved in~\cite{Dolbeault2009a,Carrillo2010a}, concavity
of $\Phi$ and a type of generalised McCann conditions are sufficient conditions to establish existence and uniqueness
of the gradient flow of the thermodynamic entropy with respect to the weighted Wasserstein metric $W_\sigma$. These
generalised $d$-dimensional McCann conditions~\cite[Definition 4.5]{Carrillo2010a} take in our context the form
\begin{equation*}
  \Phi(\rho)\Phi '(\rho)\geq (1-1/d)\int_0^\rho\Phi'(r)^2\df r.
\end{equation*}
We note that in dimension $d=1$ the McCann condition is equivalent to $\Phi'(\rho)\geq 0$ for all $\rho\geq 0$, which
is true for all mean jump rate functions $\Phi$ of zero range processes.

\begin{figure}
  \centering
  \includegraphics*[scale=.5]{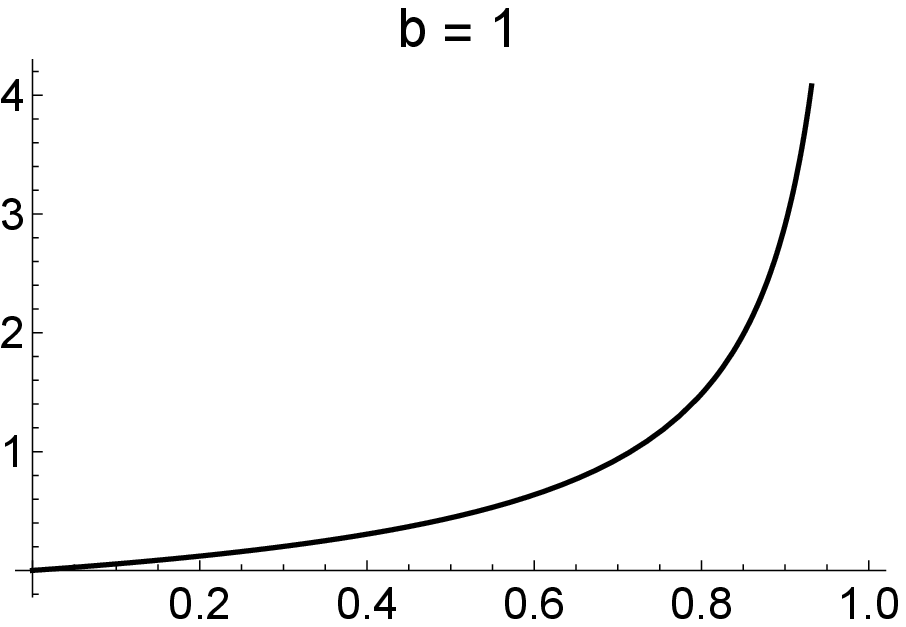}\qquad
  \includegraphics*[scale=.5]{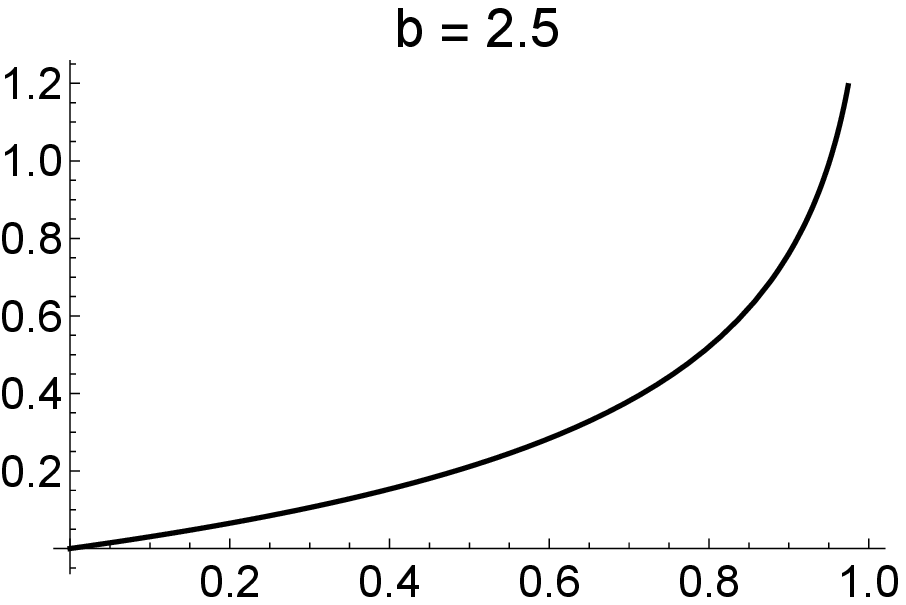}\qquad
  \includegraphics*[scale=.5]{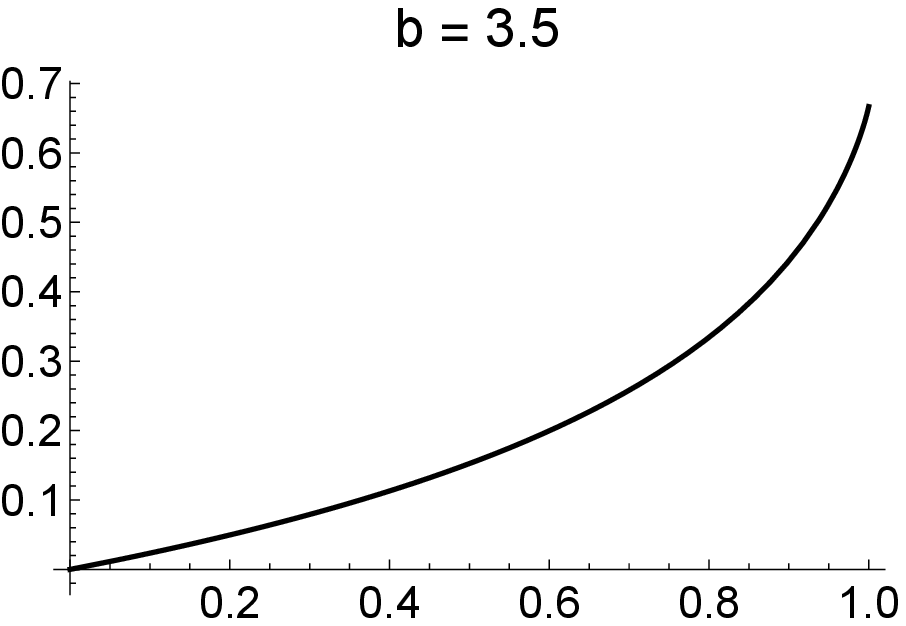}
  \caption{Graph of the density function $R_b=\Phi_b^{-1}$ in the Evans model for $b=1$, $b=2.5$ and $b=3.5$}
 \label{fig:evans}
\end{figure}

Let us also note that not all mean jump rate functions of zero range processes are concave. Two simple examples where
it is concave are $\Phi(\rho)=\rho$ and $\Phi(\rho)=\rho/(\rho+1)$, which correspond to $g(k)=k$ and $g(k)\equiv 1$,
respectively. Although the latter jump rate does not satisfy the superlinearity
assumption~\eqref{ineq:superlinjumprate}, one can consider the jump rate $g_\epsilon(k)=g(k)+\epsilon k$ for some
$\epsilon>0$, which according to~\eqref{eq:MeanJRFunction} gives the mean jump rate
$\Phi_\epsilon(\rho)=\Phi(\rho)+\epsilon\rho$. More generally, as one can see by plots the mean jump rate function
corresponding to the \emph{Evans model}
\begin{equation*}
  g_b(k)=\Big(1+\frac{b}{k}\Big)\mathbbm{1}_{\{k\geq 1\}}
\end{equation*}
is concave for all parameters $b\geq 0$. The density function $R_b=\Phi_b^{-1}$ in the Evans model is given
by~\cite[Eq. (4.5)]{Grosskinsky2003a}
\begin{equation*}
  R_b(\varphi)=\frac{\varphi{}_2F_1(2,2,2+b,\varphi)}{(1+b){}_2F_1(1,1,1+b,\varphi)},\quad 0\leq\varphi,
\end{equation*}
where ${}_2F_1(a,b,c;z):=\sum_{k=0}^\infty\frac{(a)_k(b)_k}{(c)_k}\frac{z^k}{k!}$ is the hypergeometric function and
\begin{equation*}
  (a)_k:=
  \begin{cases}
    a(a+1)\cdot\ldots\cdot(a+k-1) &\text{ if }k\geq 1\\
    1 &\text{ if }k=0
  \end{cases},
\end{equation*}
is the rising Pochhammer symbol. Fig.~\ref{fig:evans} gives plots for $R_b$ for some values of $b$.

An example of a zero-range process with non-convex density function is given by the \emph{Landim jump rate} function
given by
\begin{equation*}
  g_b(k)=
  \begin{cases}
    k &\text{ if }k=0,1\\
    (\frac{k}{k-1})^b &\text{ if }k\geq 2
  \end{cases}.
\end{equation*}
In this case, $g_b!(k)=k^b$ and the partition function is
$Z_b(\varphi)=1+\sum_{k=1}^\infty\frac{\varphi^k}{k^b}=:1+{\rm{Li}}_b(\varphi)$, where ${\rm{Li}}_b$ is the
polylogarithmic function. This results in the density function
\begin{equation*}
  R_b(\varphi)=\frac{{\rm{Li}}_{b-1}(\varphi)}{1+{\rm{Li}}_b(\varphi)}.
\end{equation*}

\begin{figure}
  \centering
  \includegraphics*[scale=.5]{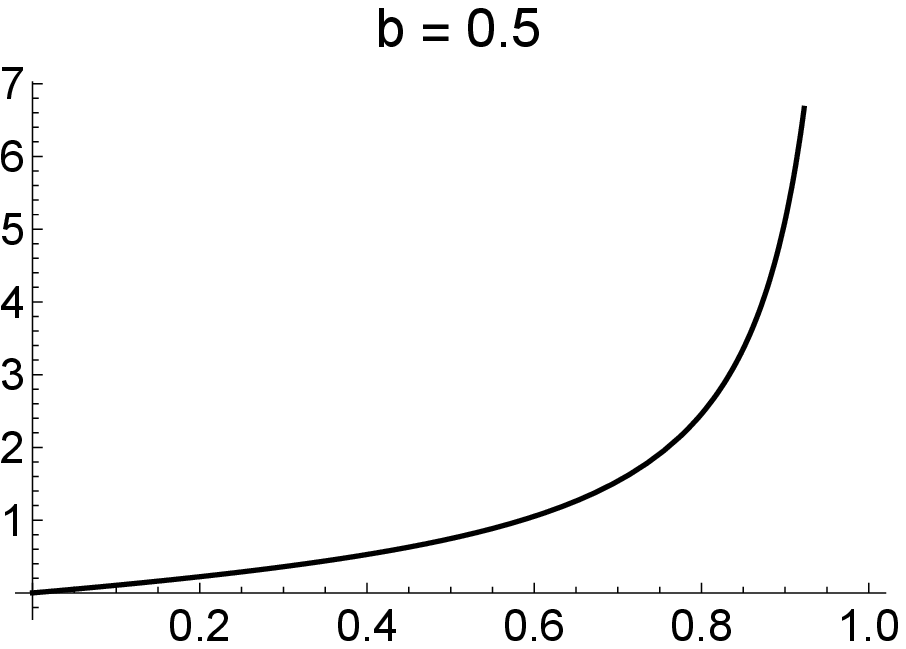}\qquad
  \includegraphics*[scale=.5]{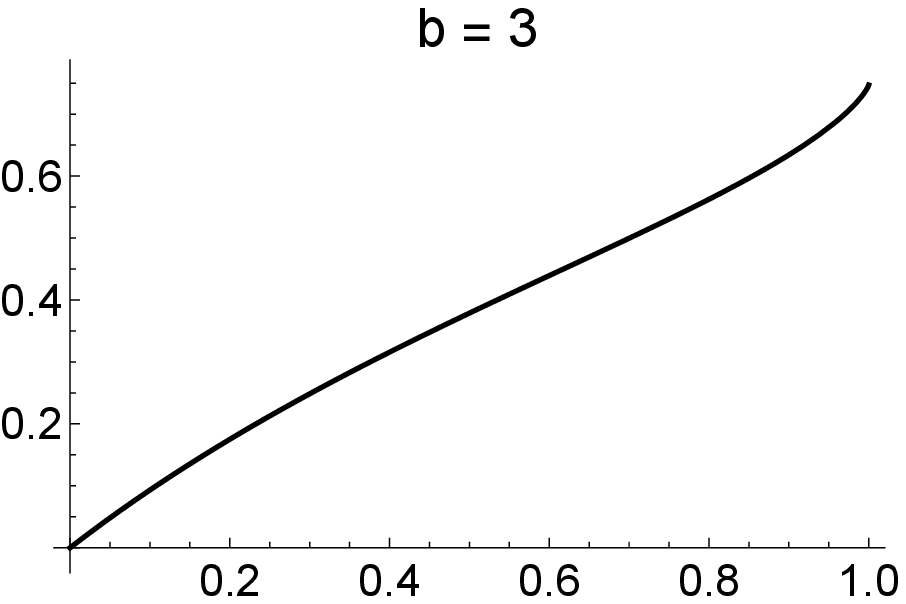}\qquad
  \includegraphics*[scale=.5]{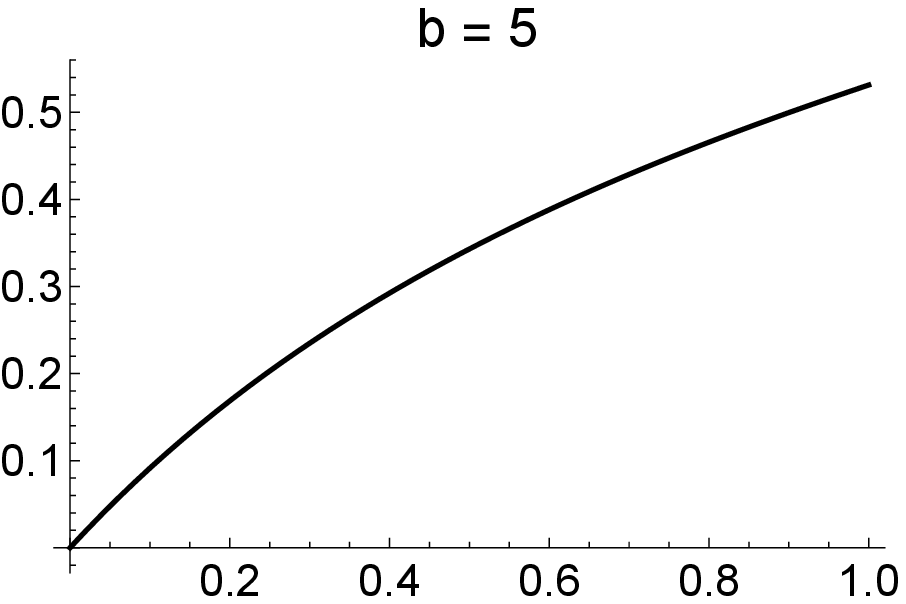}\\
  \caption{Graph of the density function $R_b=\Phi_b^{-1}$ in the Landim model for $b=0.5$, $b=3$ and $b=5$}
 \label{fig:polylog}
\end{figure}

Fig.~\ref{fig:polylog} indicates that $R_b$ is convex for small $b>0$, concave for large $b>0$ and neither of the two
for intermediate values. We close this section with the rather obvious remark that equations of porous medium type,
$\Phi(\rho) = \rho^m$, are not covered by the setting described here, as for the zero-range process $\Phi'(0)$ is
strictly positive.

\section{Connection to large deviation theory and fluctuations}
\label{sec:Connection-to-large}

\subsection{Large deviation theory}
\label{sec:Large-devi-theory}

In Section~\ref{sec:Therm-form-nonl}, we used the thermodynamic entropy $\S$ from Section~\ref{sec:Particle-models} and
used it to identify the associated (weighted Wasserstein) metric~\eqref{eq:metric}; indeed, the calculation in
section~\ref{sec:Therm-form-nonl} shows that this metric is associated to the operator
${\rm div}(\Phi(\rho)\grad\cdot)$ in~\eqref{eq:nonl-ww} which gives the gradient flow formulation of the thermodynamic
entropy. We now describe an alternative link between the zero-range process and this metric. The tool is Large
Deviation Theory, which we sketch first for a toy model. This theory is a step beyond a central limit hydrodynamic
limit theorem; on the most superficial level, large deviation theory describes the most likely event of improbable
events in terms of an exponential (un-)likelihood. As explained with the toy model, the probability of stochastic
events of $x^\epsilon$ is determined to be of the form
${\mathbb P}(\{x^\epsilon(t)\sim x(t)\}) \sim \exp\left({-\frac{1}{\epsilon} I(x(t))}\right)$; here $\sim$ means
asymptotic equivalence (precise formulations are given in any textbook on Large deviation Theory, for
example~\cite{Hollander2000a}). The functional $I$ is called the \emph{rate functional} and the central finding of this
section is that the rate functional for the zero-range process has as minimiser the nonlinear diffusion
equation~\eqref{eq:nonlin}, which appears in the norm defined by the weighted Wasserstein metric~\eqref{eq:metric}.

We first explain this link using a toy model of a stochastically perturbed one-dimensional gradient flow,
\begin{equation*}
  dX^\eps=-V'(X^\eps)dt +\sqrt{\eps}dW,
\end{equation*}
where $dW$ denotes a Wiener process. Heuristically, one expects that for small noise, $\eps$ small, trajectories will
tend to be close to the deterministic trajectory solving $\dot x = - V'(x)$.  Large deviation theory describes the
probability of observing a realisation, here a trajectory, by means of a \emph{rate functional} $I$; the probability of
finding a given trajectory is essentially $\exp(- 1/\epsilon\ I)$ as sketched now. For the toy model, the large
deviations rate functional depends on the final time $T$ and is denoted $I_T$; it is
\begin{align}
  \label{eq:LDFunctional}
  I_T(\dot x,x)&=\int_0^T\frac{1}{2}(\dot x + V'(x))^2 \df s\\
  \label{eq:LDFunctionalSplit}
   &=\int_0^t \frac{1}{2} \abs{\dot x}^2 \df s+\frac{1}{2}\int_0^T(V'(x))^2\df s +V(x(T))-V(x(0))
\end{align}
with $\dot x$ denoting the time derivative of $x$. In essence, this means that
\begin{equation*}
  {\mathbb P}(\{x^\epsilon(0)\sim x,\ x^\epsilon(T)\sim y\})
  \sim \exp\left({-\frac{1}{\epsilon}\inf I_T(\dot x,x)}\right),  
\end{equation*}
where the infimum is taken over paths $x = x(s)$ with $x(0)=x$ and $x(T)=y$. In order to find the most likely position
at $t=T,$ we have to minimise first over all paths such that $x(0)=x$ and $x(T)=y$ and then over all terminal values
$x(T)$.

Note that the first integral in~\eqref{eq:LDFunctionalSplit} is bounded from below by the distance $d(x(0),x(T))$
between the initial and the final point. Choosing $T=h\ll 1$ and re-scaling, we see that we have to minimise
\begin{equation*}
  \frac{1}{2h}d^2(x(0),x(h)) +V(x(h))+\frac h 2 \int_0^1(V'(x(sh)))^2 \df s. 
\end{equation*}

If $d(x(0),x(h))\gg h,$ then the last term is clearly of lower order, while if $d(x(0),x(h))=O(h)$ and $V$ is smooth
enough, then the last integral equals
\begin{equation*}
  \frac h 2 \int_0^1(V'(x(sh))^2 \df s= \frac h 2 V'(x(0))+O(h^2),
\end{equation*}
so again the term relevant for the minimisation is of lower order. Formally this means that we can find the most likely
position $x(h)$ by minimising the functional~\cite{Jordan1998a}
\begin{equation}
  \label{eq:JKO for beginners}
  \frac{1}{2h}d^2(x(0),x(h))+V(x(h)).
\end{equation}
The analogous computation starting with the large deviation rate functional for the empirical measure of $n$ Brownian
particles (with $n$ playing the role of the large parameter $1/\epsilon$ in the toy model; here the rate functional is
defined on function spaces)~\cite{Dawson1987a}, then one obtains an equation with the same structure as~\eqref{eq:JKO
  for beginners}, with $d$ being the Wasserstein norm, and the entropy $S$ in place of $V$ (see~\cite{Duong2013b} for
the proof). Thus, one obtains the celebrated Jordan-Kinderlehrer-Otto functional~\cite{Jordan1998a} as expansion of the
associated large deviation rate functional.

We now apply this reasoning in the context of nonlinear diffusion and more specifically the zero range process. The
large deviation rate function for the zero range process is of the form
\begin{equation}
  \label{eq:LDzerorange}
  \int_0^T\norm{\partial _t\rho(s) -\Delta \Phi(\rho(s))}_{H^{-1}_{\Phi(\rho(s))}}^2\df s
\end{equation}
(see~\cite{Koukkous2000a} and references therein); here the weighted negative Sobolev norm is defined as follows. Let
for some non-negative bounded function $w$
\begin{equation*}
  {\ip u v}_{H^1_w} :=\int _{\T^d} w \grad u \cdot \grad v \df x,
\end{equation*}
and denote by $\Delta _w^{-1}u$ the solution of
\begin{equation}
  \label{eq:negLaplace}
  \Div(w\grad (\Delta _w^{-1}u))=u.
\end{equation}
Then 
\begin{equation*}
  {\ip u v}_{H^{-1}_w}= {\ip{\Delta _w^{-1}u}{\Delta _w^{-1}v}}_{H^{1}_w}= - \int_{\T^d} u \Delta _w^{-1}v \df x,
\end{equation*}
where the last identity used integration by parts and~\eqref{eq:negLaplace}.

As seen from the reasoning above, by expanding the square in~\eqref{eq:LDzerorange}, we obtain a metric term
\begin{equation*} \int_{\T^d} \Phi \abs{\grad (\Delta _{\Phi(\rho_t)}^{-1}\partial_t \rho )}^2 \df x
  =g_\sigma(\partial_t\rho_t,\partial_t\rho_t),
\end{equation*}
where $g$ is as in~\eqref{eq:metric}. This follows since if $v_\zeta=\sigma\nabla\xi_\zeta$ is the vector field
representing a tangent vector $\zeta$ at $\rho$ via~\eqref{eq:eta}, then
$\zeta=-\Div(\rho v_\zeta)=-\Div(\rho\sigma\nabla\xi_\zeta)=-\Div\big(\Phi(\rho)\nabla\xi_\zeta\big)$ and thus
$\Delta_{\Phi(\rho)}^{-1}(\zeta)=\xi_\zeta$.

The null Lagrangian (the mixed term) equals
\begin{equation*}
  \int_0^T\int _{\T^d} \partial_t\rho\Delta_{\Phi(\rho)}^{-1}(\Delta \Phi(\rho)) \df x \df t,
\end{equation*}
so it remains to compute $\Delta_{\Phi(\rho)}^{-1}(\Delta \Phi(\rho))$.  As $\log \Phi(\rho)$ solves
\begin{equation*}
  \Div (\Phi(\rho)\nabla \log \Phi(\rho))=\Delta \Phi(\rho),
\end{equation*}
we have $\Delta_{\Phi(\rho)}^{-1}(\Delta \Phi(\rho))=\log (\Phi(\rho))$, so if $S'(\rho)=\log(\Phi(\rho))$ as
in~\eqref{eq:thermo-ent-diff}, we obtain
\begin{equation}\label{largedev1}
  \int_0^T\int _{\T^d} \partial_t\rho\Delta_\Phi^{-1}(\Delta \Phi(\rho)) \df x \df t=
  \int_{\T^d} S(\rho(T)) \df x-\int_{\T^d} S(\rho(0)) \df x,
\end{equation}
and recover the thermodynamic entropy $\S$ of~\eqref{eq:entropy}.

The remaining quadratic term is for small time intervals formally of lower order. Thus in summary, an expansion of the
large deviation rate functional~\eqref{eq:LDzerorange} for nonlinear diffusion associated with the zero range process
yields a variational formulation of Jordan-Otto-Kinderlehrer type, with the metric being the weighted Wasserstein
metric considered in this article, and the functional being the thermodynamic entropy.

\subsection{Fluctuating hydrodynamics}
\label{sec:Fluct-hydr}

The previous Subsection~\ref{sec:Large-devi-theory} showed that the thermodynamic metric is related not only to the
limit equation, but also to the probability of deviations from it. At least at a formal level, this connection is even
stronger, as we now explain. Formally, the metric is given by a symmetric linear operator, which has an inverse that
can be interpreted as covariance matrix. If we add a suitable space-time noise to the limit partial differential
equation, then it formally has not only the same large deviations rate functionals, but finite moments of the
fluctuations (that is, the difference between stochastic process and deterministic limit) converge to the same process
as for the original particle model. This is now explained in more detail. Consider the equations
\begin{align}
  d\rho^\eps&=\Delta (\Phi(\rho^\eps)) dt+\Div(\sqrt{\eps \sigma(\rho^\eps)\rho^\eps}\dot W)  \label{eq:SPDE-a}\\ 
  \partial_t\rho&=\Delta(\Phi(\rho)), \label{eq:SPDE-b}
\end{align}
where $\dot W$ is space-time white noise.  We point out that the stochastic PDE (SPDE)~\eqref{eq:SPDE-a} is not
expected to have strong (i.e., pathwise) solutions. To our knowledge, the suitable notion of generalised solutions is
an open problem. Consider, however, the fluctuations, i.e., the differences between solutions of~\eqref{eq:SPDE-a} and
the deterministic PDE~\eqref{eq:SPDE-b}, re-scaled with the scaling of the central limit theorem.  As
\begin{equation*}
  Z^\eps:=\rho-\rho^\eps
\end{equation*}
is small, we can linearise around $\rho $ and obtain formally that a limit (in law) $Z^0$ of $\eps^{-1/2}Z^\eps$ should
solve the \emph{linear} SPDE
\begin{equation}
  \label{fluctuation}
  dZ^0=\Delta( \Phi'(\rho)Z^0)dt+\Div(\sqrt{ \sigma(\rho^\eps)\rho^\eps}\dot W),
\end{equation}
i.e., it is an Ornstein-Uhlenbeck process. Here $\Phi'$ is the \emph{self-diffusion} coefficient of the zero range
process, while $\sigma$ is the \emph{bulk-diffusion} coefficient.
 
This convergence statement for the fluctuations can be stated in the following weaker form. Consider for smooth
$\g\colon\R^d\to \R$ the process
\begin{equation*}
  Y^\eps:=\eps^{-1/2}\ip{\g}{\rho-\rho^\eps},
\end{equation*}
then by~\eqref{fluctuation}, an application of Ito's formula (see, e.g.,~\cite[Chapter 4.5]{Da-Prato2014a}), and the
(formal) limit $\eps\to 0$ we get for any smooth $F\colon \R\to \R$ that
\begin{multline}
  \label{martingaleproblem}
  F(Y^0(t))-F(Y^0(0))-\int_0^tF'(Y^0(s)) \ip{Y^0(s)}{\Phi'(\rho(s)) \Delta \g} ds\\
  -\frac{1}{2} \int_0^t
  F''(Y^0(s))\ip{\sigma(\rho(s)) \rho(s) \grad \g}{\grad \g} ds 
\end{multline}
is a \emph{martingale}. 

Note that in the last term, the expression $\ip{\sigma(\rho(s)) \rho(s) \grad \g}{\grad \g}$ can be written as
$\norm{\g}_{H^1_\Phi(\rho)}^2$, which implies that the large deviations should be related to the metric with the inverse
operator, i.e., $\norm{\g}_{H^{-1}_\Phi(\rho)}^2$, as found in the previous subsection.

We now relate the martingale~\eqref{martingaleproblem} to the zero range process. We consider the empirical measure of
the zero range process, i.e., for smooth $\g$
\begin{equation*}
  \ip{\rho^N}{\g}:=\frac{1}{N^d}\sum_{x\in\T_N^d}\g\left(\frac{x}{N}\right)\eta(x)
\end{equation*}
and the fluctuations 
\begin{equation*}
  Y^N:=N^{-d/2}\ip{\g}{\rho-\rho^N}.
\end{equation*}
It can be shown~\cite{Ferrari1988a,Gielis1998a,Jankowski2006a} for a large class of zero range processes that for any
smooth test function, $F(Y^N)$ converge to a solution of the martingale problem~\eqref{martingaleproblem}. In other
words, the limit of the fluctuations of the zero range process is in law equal to those of the
SPDE~\eqref{eq:SPDE-a}. This can be seen as justification of the complicated noise term in~\eqref{eq:SPDE-a}; precisely
this noise term is related to fluctuations of the zero range process via the martingale
expression~\eqref{martingaleproblem}. We remark that~\eqref{eq:SPDE-a} is of fluctuating hydrodynamics type. In the
linear case $\Phi(\rho) = \rho$, the multiplicative conservative noise in~\eqref{eq:SPDE-a} has been derived by
Dean~\cite{Dean1996a}, and using quite different lines of arguments by Sturm and von Renesse~\cite{Renesse2009a}; the
link between the noise and the geometry has been explained by a formal large deviation argument~\cite{Jack2014a}.

While the discussion in this paper is limited to a sub-class of zero range processes, it therefore does not cover all
nonlinearities $\Phi(\rho)$ but only those that occur as hydrodynamic limit of such a process.  Yet, we remark that a
solution of \eqref{eq:SPDE-a} for a certain range of $\epsilon>0$ would formally yield a stochastic process whose
hydrodynamic limit is a gradient flow of the entropy defined by \eqref{eq:thermo-ent} with respect to the weighted
Wasserstein metric, and whose large deviation rate functional is as in Subsection~\ref{sec:Large-devi-theory}.  The
existence of solutions to \eqref{eq:SPDE-a}, however, is to our knowledge in general an open problem.

\paragraph{Acknowledgements} We thank Mark A.~Peletier for helpful and stimulating discussions. All authors gratefully
acknowledge funding from the Leverhulme Trust, RPG-2013-261, and JZ was partially funded by the EPSRC (EP/K027743/1)
and a Royal Society Wolfson Research Merit Award. The authors thank Marcus Kaiser and an anonymous reviewer for
valuable comments and suggestions.


\def\cprime{$'$} \def\cprime{$'$} \def\cprime{$'$}
  \def\polhk#1{\setbox0=\hbox{#1}{\ooalign{\hidewidth
  \lower1.5ex\hbox{`}\hidewidth\crcr\unhbox0}}} \def\cprime{$'$}
  \def\cprime{$'$}

\end{document}